\documentclass[aps,prb,a4paper,10pt,twocolumn,showpacs,floatfix,superscriptaddress,preprintnumbers,longbibliography]{revtex4-2}
\usepackage{float}
\usepackage{dcolumn,graphicx,color,booktabs,microtype,afterpage}
\usepackage{amssymb}
\usepackage{amsmath}
\usepackage[charter,greekuppercase=italicized]{mathdesign}
\usepackage{sidecap}
\usepackage[mathlines]{lineno}

\graphicspath{{./}{figure/}}
\renewcommand{\tablename}{Table}
\makeatletter\renewcommand{\fnum@figure}[1]{\figurename~\thefigure.~}\makeatother
\makeatletter\renewcommand{\fnum@table}[1]{\tablename~\thetable.}\makeatother

\newcount\hh \newcount\mm
\hh=\time \divide\hh by 60
\mm=\hh \multiply\mm by 60 \mm=-\mm
\advance\mm by \time
\def\now{\number\hh:\ifnum\mm<10{}0\fi\number\mm}

\usepackage[colorlinks,plainpages=false,linkcolor=blue,urlcolor=blue,citecolor=blue,pdfpagemode=UseNone,pdfstartview=FitBH]{hyperref}
\usepackage{nicefrac}

\newcommand{\tcr}[1]{\textcolor{black}{#1}}

\begin{document}

\makeatletter\renewcommand{\ps@plain}{%
\def\@evenhead{\hfill\itshape\rightmark}%
\def\@oddhead{\itshape\leftmark\hfill}%
\renewcommand{\@evenfoot}{\hfill\small{--~\thepage~--}\hfill}%
\renewcommand{\@oddfoot}{\hfill\small{--~\thepage~--}\hfill}%
}\makeatother\pagestyle{plain}

\preprint{\textit{Preprint: \today, \now}} 
\title{Evidence of unconventional pairing in the quasi two-dimensional \\ CuIr$_2$Te$_4$ superconductor}

%
\author{T.\ Shang}\email[Corresponding author:\\]{tshang@phy.ecnu.edu.cn}\thanks{These authors contributed equally}
\affiliation{Key Laboratory of Polar Materials and Devices (MOE), School of Physics and Electronic Science, East China Normal University, Shanghai 200241, China}
%
\author{ Y.\ Chen}\thanks{These authors contributed equally}
\affiliation{Center for Correlated Matter and Department of Physics, Zhejiang University, Hangzhou 310058, China}

\author{W.\ Xie}
\affiliation{Institute of High Energy Physics, Chinese Academy of Sciences, Beijing 100049, China} 
\affiliation{Spallation Neutron Source Science Center, Dongguan 523803, China}
%
%
%
%
%
%
\author{D.\ J.\ Gawryluk}
\affiliation{Laboratory for Multiscale Materials Experiments, Paul Scherrer Institut, CH-5232 Villigen PSI, Switzerland}
\author{R.\ Gupta}
\affiliation{Laboratory for Muon-Spin Spectroscopy, Paul Scherrer Institut, CH-5232 Villigen PSI, Switzerland}
\author{R.\ Khasanov}
\affiliation{Laboratory for Muon-Spin Spectroscopy, Paul Scherrer  Institut, CH-5232 Villigen PSI, Switzerland}
\author{X.\ Y.\ Zhu}
\affiliation{Key Laboratory of Polar Materials and Devices (MOE), School of Physics and Electronic Science, East China Normal University, Shanghai 200241, China}
\author{H.\ Zhang}
\affiliation{Key Laboratory of Polar Materials and Devices (MOE), School of Physics and Electronic Science, East China Normal University, Shanghai 200241, China}
\author{Z.~X.~Zhen}
\affiliation{Key Laboratory of Polar Materials and Devices (MOE), School of Physics and Electronic Science, East China Normal University, Shanghai 200241, China}
\author{B.\ C.\ Yu}
\affiliation{Key Laboratory of Polar Materials and Devices (MOE), School of Physics and Electronic Science, East China Normal University, Shanghai 200241, China}
\author{Z.\ Zhou}
\affiliation{Key Laboratory of Nanophotonic Materials and Devices \&
Key Laboratory of Nanodevices and Applications,
Suzhou Institute of Nano-Tech and Nano-Bionics (SINANO), CAS, Suzhou 215123, China}

\author{Y.\ Xu}
\affiliation{Key Laboratory of Polar Materials and Devices (MOE), School of Physics and Electronic Science, East China Normal University, Shanghai 200241, China}
\author{Q.\ F.\ Zhan}
\affiliation{Key Laboratory of Polar Materials and Devices (MOE), School of Physics and Electronic Science, East China Normal University, Shanghai 200241, China}
\author{E.\ Pomjakushina}
\affiliation{Laboratory for Multiscale Materials Experiments, Paul Scherrer Institut, CH-5232 Villigen PSI, Switzerland}
\author{H.\ Q.\ Yuan}
\affiliation{Center for Correlated Matter and Department of Physics, Zhejiang University, Hangzhou 310058, China}

%
\author{T.\ Shiroka}
\affiliation{Laboratory for Muon-Spin Spectroscopy, Paul Scherrer Institut, CH-5232 Villigen PSI, Switzerland}
\affiliation{Laboratorium f\"ur Festk\"orperphysik, ETH Z\"urich, CH-8093 Z\"urich, Switzerland}
\begin{abstract}
The CuIr$_{2-x}$Ru$_x$Te$_4$ superconductors (with a $T_c$ around 2.8\,K)
can host charge-density waves, whose onset and interplay with
superconductivity are not well known at a microscopic level.
Here, we report a comprehensive study of the $x$ = 0 and 0.05 cases, whose 
superconductivity was characterized via 
electrical-resistivity-, magnetization-, and heat-capacity measurements,
while their microscopic superconducting properties were studied via
muon-spin rotation and relaxation ($\mu$SR).
In CuIr$_{2-x}$Ru$_x$Te$_4$, both the temperature-dependent electronic
specific heat and the superfluid density (determined via transverse-field
$\mu$SR) are best described by a two-gap ($s+d$)-wave model, comprising
a nodeless gap and a gap with nodes.  
The multigap superconductivity is also
supported by the temperature
dependence of the upper critical field $H_\mathrm{c2}(T)$.
However, under applied pressure, a charge-density-wave order
starts to develop and, as a consequence, the superconductivity of
CuIr$_2$Te$_4$ achieves a more conventional $s$-wave character.
From a series of experiments, 
we provide ample evidence that the CuIr$_{2-x}$Ru$_x$Te$_4$ family belongs to the rare cases, where an unconventional superconducting
pairing is found near a charge-density-wave quantum critical point. 
\end{abstract}

\maketitle\enlargethispage{3pt}

\vspace{-5pt}
\section{\label{sec:Introduction}Introduction}\enlargethispage{8pt}
The interplay between 
different electronic ground states is one of
the fundamental topics in current condensed-matter physics. 
Notably, the materials exhibiting high\--tem\-pe\-ra\-ture- or
unconventional superconductivity (SC), as e.g., heavy fermions,
cuprates, or iron-based superconductors~\cite{Monthoux2007,Mazin2010,Keimer2015,Fernandes2014},
are particularly relevant in this respect since, in most of them,
the different types of order are closely related 
or even competing. Materials which sustain a charge-density-wave (CDW)
order are renowned as suitable
systems for investigating the coexistence
and interplay between these different ground states~\cite{Morosan2006,Cho2018,Ortuz2020,Jiang2021,Nie2022}.
The $AT_2X_4$ chalcogenides (with $A$, $T$ = transition metals, and $X$ = O, S, Se, Te) belong to this class and exhibit varied crystal structures with intriguing electronic 
properties. In particular, the Cu$T_2X_4$ family has attracted special attention, due to its multifaceted ground states, including also
SC and CDW. 
For instance, CuIr$_2$S$_4$ undergoes a metal-to-insulator transition at 230\,K~\cite{Nagata1994,Furubayashi1994,Nagata1998}, which is suppressed via Cu/Zn substitution 
to give rise to a dome-like SC phase (with maximum $T_c$ = 3.4\,K)~\cite{Suzuki1999}. 
Similarly, in the CuIr$_2$Se$_4$ case, superconductivity with $T_c = 1.76$\,K can be induced via Ir/Pt substitution~\cite{Luo2013}.
Further, CuV$_2$S$_4$ is known to exhibit three different CDW transitions (between 55 and 90\,K), before it enters the SC phase at $T_c$ = 4.4\,K~\cite{Fleming1981}. 
Although hundreds of $AT_2X_4$-type materials have been discovered and examined, superconductivity has only been found in Cu-based sulpho- or seleno-spinels. 
An exception to this are telluride spinels (yet another chalcogen), which adopt lower dimensional crystal structures compared to the S- or Se based-spinels, with
CuIr$_2$Te$_4$ recently shown to display SC~\cite{Yan2019}.
Since, unlike its preceding group-16 elements, Te is a metalloid, we expect its properties to differ from those of  S- or Se cases.

\begin{figure}[!thp]
	\centering
	\vspace{-1ex}%
	\includegraphics[width=0.45\textwidth,angle=0]{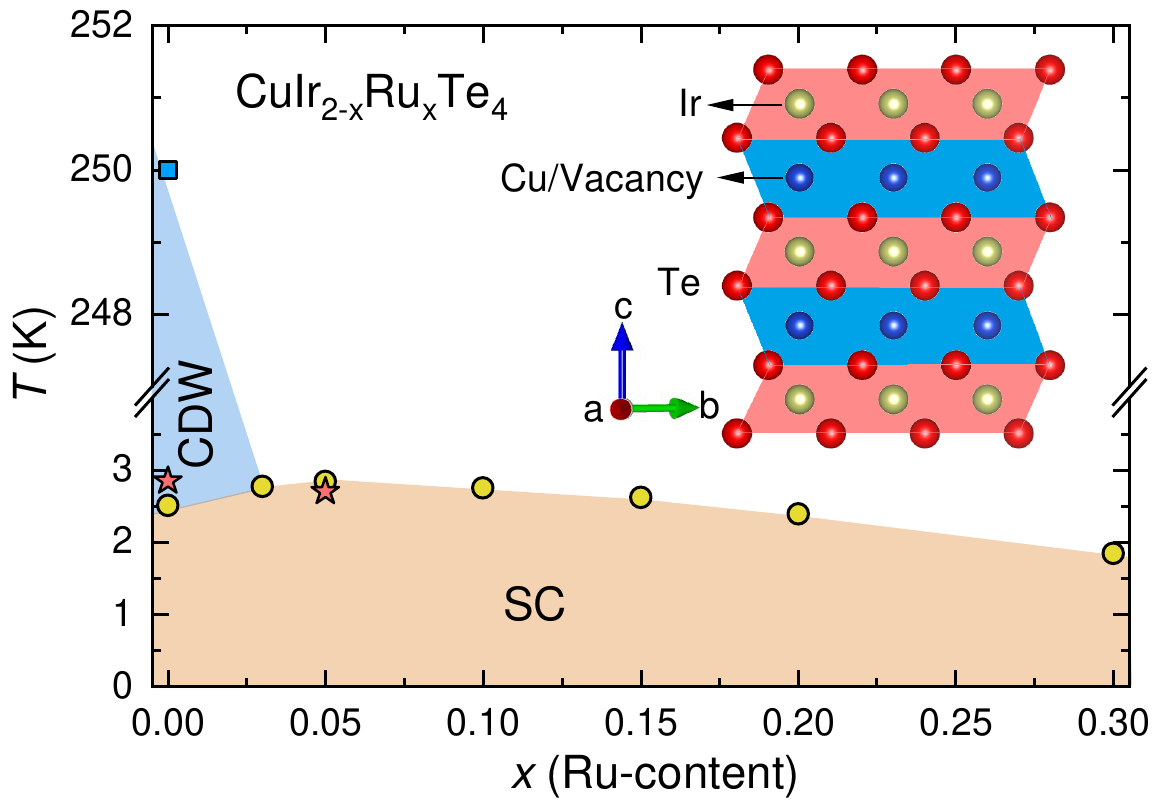}
	\caption{\label{fig:phase_digram} \tcr{Phase diagram of CuIr$_{2-x}$Ru$_x$Te$_4$.
		The stars refer to the current work, while the other symbols represent
		data taken from Ref.~\onlinecite{Dong2019}. 
		The CDW transition temperature $T_\mathrm{CDW}$ was determined from
		the temperature-dependent electrical resistivity, while the
		superconducting transition temperatures $T_c$ were derived from magnetic-susceptibility measurements.  
		The inset shows the crystal structure of CuIr$_2$Te$_4$
        (i.e., without depicting the Cu vacancies) viewed
		along the [100]-direction, clearly demonstrating its quasi-two-dimensional character.
		Blue-, yellow-, and red spheres represent Cu-, Ir-, and Te atoms, respectively.}} 
\end{figure}

\tcr{CuIr$_2$Te$_4$ crystallizes in a disordered trigonal structure with
space group $P\overline{3}m1$ (No.~164), where the Cu atoms and vacancies
are randomly distributed in the Cu layers. 
CuIr$_2$Te$_4$ exhibits a quasi-two-dimensional crystal structure,
where IrTe$_2$ layers are intercalated by Cu planes (see inset in Fig.~\ref{fig:phase_digram}).}
Such ``sandwich''-like structure is also
encountered in Cu$_x$Bi$_2$Se$_3$~\cite{Hor2010}, a prime example 
of topological superconductor~\cite{Sasaki2011}. CuIr$_2$Te$_4$ undergoes a first-order CDW transition around 250\,K, where both the electrical resistivity and magnetization exhibit
clear anomalies, with a significant temperature hysteresis~\cite{Yan2019,Nagata1999}. More interestingly, by further decreasing the temperature below $T_c$ = 2.5\,K, CuIr$_2$Te$_4$ becomes a superconductor~\cite{Yan2019}. 
Upon substituting Ir with Ru, in CuIr$_{2-x}$Ru$_x$Te$_4$, the CDW order is
quickly suppressed at $x$ = 0.03, while the superconducting transition temperature
increases up to 2.8\,K (for $x$ = 0.05, see Fig.~\ref{fig:phase_digram})~\cite{Dong2019}.
\tcr{Similar features have been found also in Al-, Ti-, and Zr-substituted
CuIr$_2$Te$_4$~\cite{Zeng2021,Yan2022,Zeng2022b}. 
Although at low doping Cr-substituted CuIr$_2$Te$_4$ samples show
similar behavior to the above families, once the Cr-content is above
0.25 (i.e., $x > 0.25$), a ferromagnetic order occurs~\cite{Zeng2022}.}
Such dome-like superconducting phase resembles that of unconventional superconductors~\cite{Monthoux2007,Mazin2010,Keimer2015,Fernandes2014} and transition-metal dichalcogenides~\cite{Morosan2006,Cho2018}.
As in them, also in CuIr$_2$Te$_4$ the dome
may signal the presence of a CDW quantum critical point (QCP), with
the increase in 
$T_c$ reflecting the enhanced quantum fluctuations near the QCP.
Upon increasing the Ru content above $x$ = 0.3, the SC is destroyed.  
According to electronic band-structure calculations, the density of states near the Fermi level consists mostly of Te-$p$ and Ir-$d$ orbitals~\cite{Yan2019}, 
both experiencing a large spin-orbit coupling, potentially leading to unconventional superconducting properties. Here, the competition between CDW and
%
%
SC is possibly tuned by modifications of the density of states and Fermi surface via chemical doping.    

Since, in CuIr$_2$Te$_4$, both CDW and SC can be easily tuned via
chemical substitution, this represents an ideal system for
investigating the interplay between the two. Although certain properties of CuIr$_{2-x}$Ru$_x$Te$_4$ family have
been examined, and electronic band-structure calculations are available, at a microscopic level, its superconducting
properties, in particular the superconducting order parameter, have not been explored and
await further investigation.

In this work, after synthesizing CuIr$_{2-x}$Ru$_x$Te$_4$ ($x = 0$ and 0.05)
samples, we systematically studied their superconducting properties by
means of electrical-resistivity-, magnetization-, and heat-capacity measurements,
complemented by muon-spin relaxation and rotation ($\mu$SR) method. 
Certain measurements were also performed under external pressure (up to 2.5\,GPa).
Both the superfluid density and the electronic specific heat of CuIr$_{2-x}$Ru$_x$Te$_4$ are best described by a two-gap model, consisting
of a nodeless gap and a gap with nodes.
The evidence of a nodal gap is our key observation, which suggests the
CuIr$_{2-x}$Ru$_x$Te$_4$ family as a remarkable system where competing orders can lead to unconventional SC behavior.

\section{Experimental details\label{sec:details}} \enlargethispage{8pt}
Polycrystalline CuIr$_{2-x}$Ru$_x$Te$_4$ samples, with $x$ = 0
and 0.05, were prepared by the solid-state reaction method (the details can be found
in Ref.~\onlinecite{Dong2019}).  
The crystal structure and phase purity were checked by powder 
x-ray diffraction, confirming the trigonal
structure of CuIr$_{2-x}$Ru$_x$Te$_4$ ($P\overline{3}m1$, No.~164).
The superconductivity was characterized by electrical-resistivity-,
heat-capacity-, and magnetization measurements, performed on a
Quantum Design physical property measurement 
system (PPMS) and a magnetic property measurement system (MPMS), respectively.
For the electrical-resistivity- and ac-susceptibility measurements under
%
\begin{figure}[!thp]
	\centering
	\includegraphics[width=0.5\textwidth,angle=0]{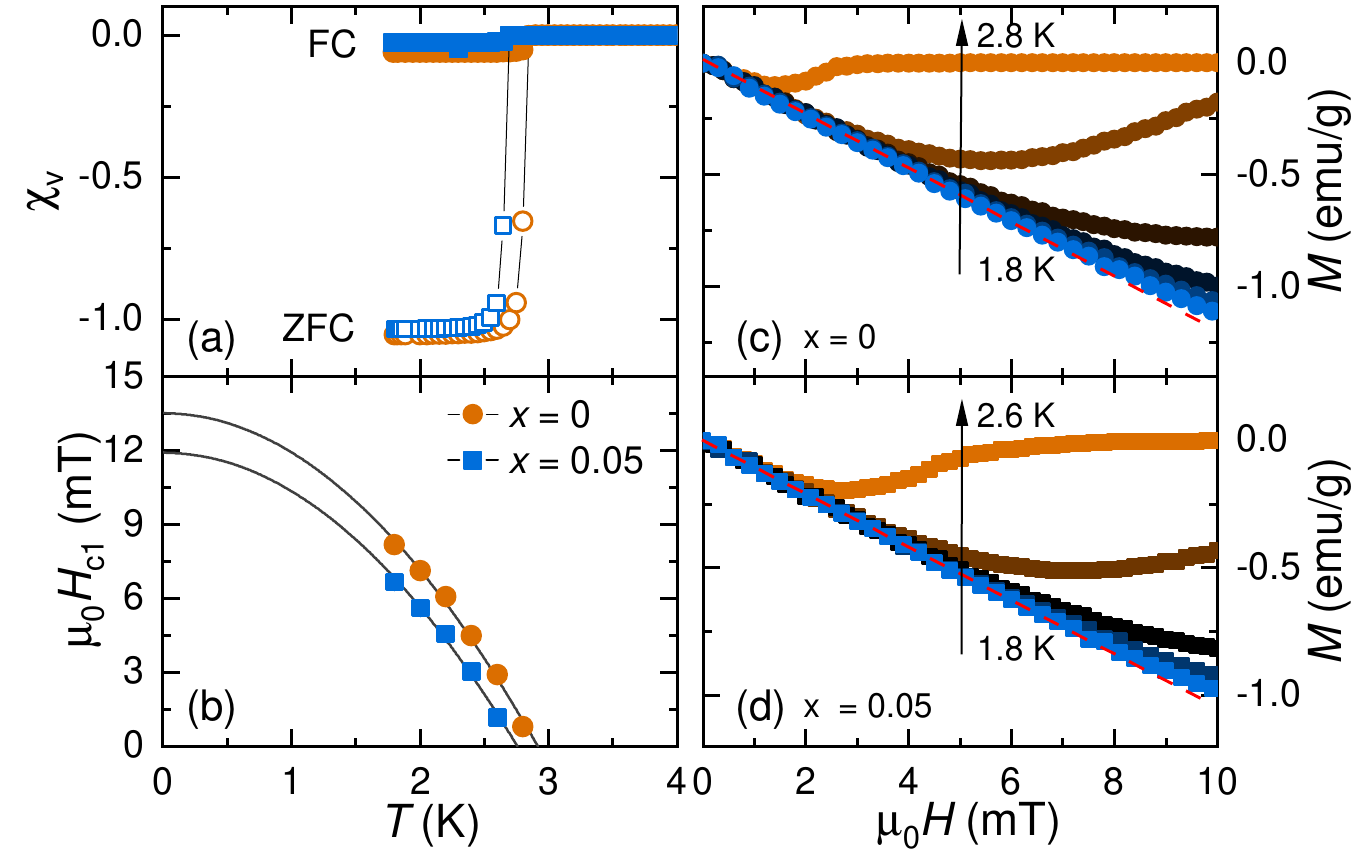}
	\caption{\label{fig:Hc1}(a) Temperature-dependent magnetic susceptibility 
		$\chi_\mathrm{V}(T)$ of CuIr$_{2-x}$Ru$_x$Te$_4$, with $x$ = 0 and 0.05. The ZFC- and FC-magnetic
		susceptibilities were measured in a field of $\mu_0$$H$ = 1\,mT. 
		(b) Lower critical fields $H_\mathrm{c1}$ vs.\ temperature. Solid lines are fits to $\mu_{0}H_\mathrm{c1}(T) =\mu_{0}H_\mathrm{c1}(0)[1-(T/T_{c})^2]$.
		Field-dependent magnetization curves
		collected at various temperatures after cooling the  CuIr$_2$Te$_4$ (c) and CuIr$_{1.95}$Ru$_{0.05}$Te$_4$ (d)
		samples in zero field. For each temperature, $H_\mathrm{c1}$ is determined as the value where $M(H)$ starts deviating from linearity [dashed lines in panels (c) and (d)].}
\end{figure}
%
pressure we employed a BeCu piston-cylinder cell, with Daphne oil 7373
used as the pressure transmitting medium.

The bulk $\mu$SR measurements were carried out at the multipurpose 
surface-muon spectrometer (Dolly) of the Swiss muon source at Paul 
Scherrer Institut, Villigen, Switzerland. 
In this study, we performed
mostly transverse-field (TF-) $\mu$SR 
measurements, which allowed us
to determine the temperature evolution of the magnetic penetration depth and thus,
of the superfluid density.  
All the $\mu$SR spectra were collected upon heating and were analyzed 
by means of the \texttt{musrfit} software package~\cite{Suter2012}.

\section{Results and discussion\label{sec:results}}\enlargethispage{8pt}
\subsection{Lower- and upper critical fields}

The bulk superconductivity of CuIr$_{2-x}$Ru$_x$Te$_4$ was
first characterized by magnetic-susceptibility measurements,
using both field-cooled (FC) and zero-field-cooled (ZFC) protocols
in an applied field of 1\,mT. As shown in Fig.~\ref{fig:Hc1}(a),
a clear diamagnetic signal appears below the superconducting transition
at $T_c$ = 2.85 and 2.7\,K for CuIr$_2$Te$_4$ and CuIr$_{1.95}$Ru$_{0.05}$Te$_4$, respectively.  
The rather sharp transitions (with a $\Delta$$T$ $\sim$ 0.1\,K,
three times smaller than in a previous work~\cite{Yan2019})
indicate a good sample quality. After accounting for the demagnetizing 
factor, the superconducting shielding fraction of both samples
is $\sim$100\%, indicative of bulk SC, as further
confirmed by heat-capacity- and $\mu$SR measurements (see below).
To determine the lower critical field $H_\mathrm{c1}$, essential for performing $\mu$SR measurements
on type-II superconductors, the field-dependent
magnetization $M(H)$ was collected at various temperatures.
The $M(H)$ curves are shown in Fig.~\ref{fig:Hc1}(c) and \ref{fig:Hc1}(d) for CuIr$_2$Te$_4$ and CuIr$_{1.95}$Ru$_{0.05}$Te$_4$, respectively. 
The estimated $H_\mathrm{c1}$ values (accounting for the demagnetization factor) as a function of temperature are summarized
in Fig.~\ref{fig:Hc1}(b). They result in $\mu_0$$H_\mathrm{c1}(0)$ = 13.5(3) and 11.9(3)\,mT for the pure and the Ru-substituted sample, respectively.

\begin{figure}[!htp]
	\centering
	\includegraphics[width=0.5\textwidth,angle= 0]{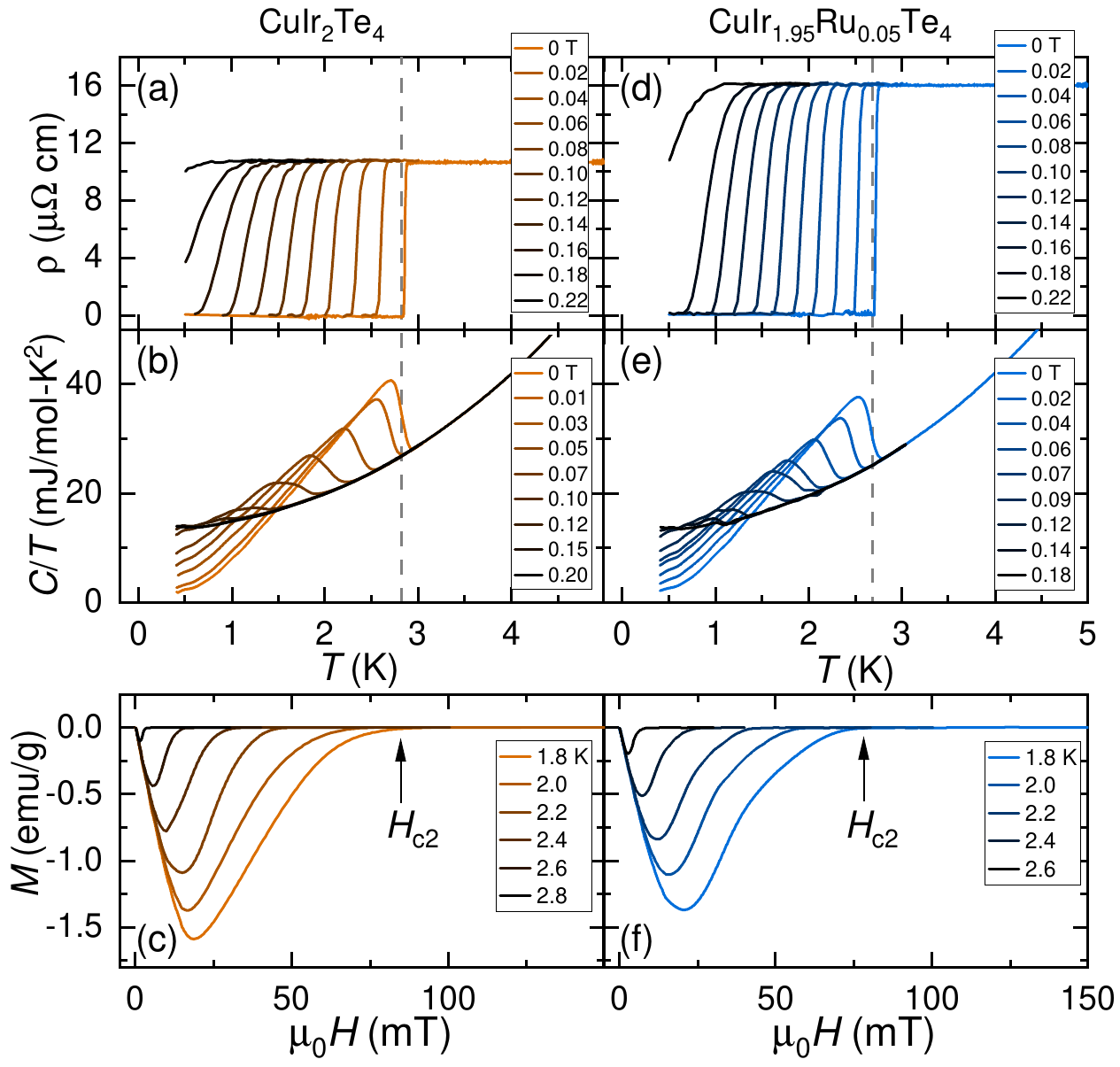}
	\caption{\label{fig:Hc2_raw} \tcr{Temperature-dependent electrical 
		resistivity $\rho(T,H)$ (a) and specific heat $C(T,H)/T$ (b) for CuIr$_2$Te$_4$ measured at
		various applied magnetic fields.  
		(c) Field-dependent magnetization $M(H,T)$ collected at various temperatures for CuIr$_2$Te$_4$. (d)-(f) The analogous results for CuIr$_{1.95}$Ru$_{0.05}$Te$_4$.
		For the $\rho(T,H)$ measurements, $T_c$  was defined as the onset of zero
		resistivity; while for the $C(T,H)/T$ measurements, $T_c$ was defined as the midpoint of superconducting transition (marked by dashed lines).
		As indicated by arrows in panels (c) and (f), $H_\mathrm{c2}$ was chosen as the field where the diamagnetic 
		response in $M(H,T)$ vanishes.}}
\end{figure}

To investigate the upper critical field $H_\mathrm{c2}$ of CuIr$_{2-x}$Ru$_x$Te$_4$,
measurements of the temperature-dependent electrical resistivity $\rho(T,H)$ and specific heat $C(T,H)/T$
at various applied magnetic fields, as well as the field-dependent 
magnetization $M(H,T)$ at various temperatures were performed. These results are summarized in Fig.~\ref{fig:Hc2_raw}(a)-(c) and \ref{fig:Hc2_raw}(d)-(f) for CuIr$_2$Te$_4$ and CuIr$_{1.95}$Ru$_{0.05}$Te$_4$, respectively. 
The distinct specific-heat jump at $T_c$ confirms again the bulk nature of SC in CuIr$_{2-x}$Ru$_x$Te$_4$.  
Upon increasing the magnetic field, the superconducting transition in $\rho(T)$ and $C(T)/T$ shifts to lower temperatures. 
In the $M(H)$ data, the diamagnetic signal vanishes once the applied magnetic field exceeds the upper critical field 
$H_\mathrm{c2}$, as indicated by the arrows in Fig.~\ref{fig:Hc2_raw}(b) and \ref{fig:Hc2_raw}(f). We found that the onset of 
zero resistivity corresponds to the midpoint of the superconducting
transition in the specific heat (indicated by dashed lines in Fig.~\ref{fig:Hc2_raw}).

\begin{figure}[!thp]
	\centering
	\includegraphics[width=0.45\textwidth,angle= 0]{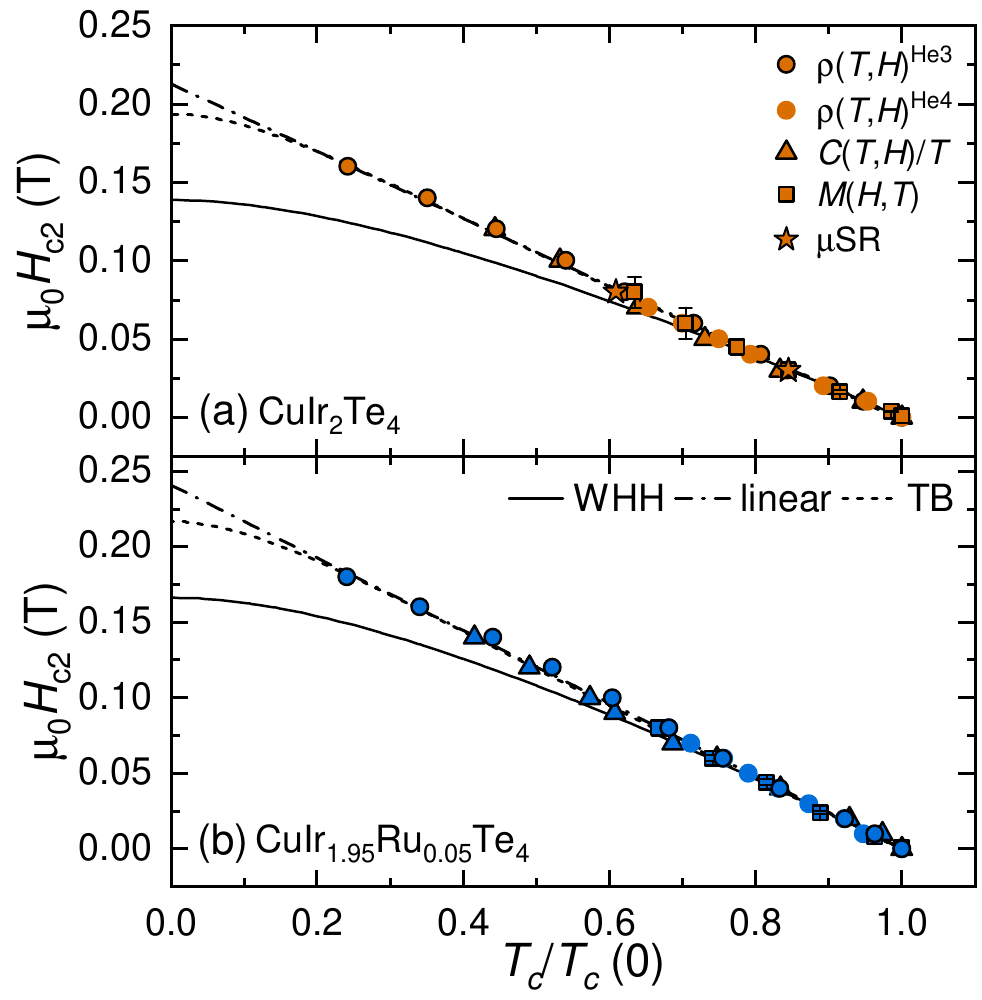}
	\caption{\label{fig:Hc2}
		Upper critical field $H_\mathrm{c2}$ vs.\ 
		the reduced superconducting transition temperature $T_c/T_c(0)$
		for CuIr$_2$Te$_4$ (a) and CuIr$_{1.95}$Ru$_{0.05}$Te$_4$ (b).   
		The $T_c$ and $H_\mathrm{c2}$ values were determined from 
		the measurements shown in Fig.~\ref{fig:Hc2_raw} and Fig.~\ref{fig:superfluid}, which are highly consistent. The temperature-dependent electrical resistivity was measured using both He-3 and He-4 cryostats. 
		The solid- and the dotted lines represent fits to the WHH and TB models, while the dash-dotted lines indicate a linear temperature dependence.}	
\end{figure}

The upper critical fields $H_\mathrm{c2}$ 
vs.\ the reduced superconducting temperatures $T_c/T_c(0)$ are summarized in 
Figs.~\ref{fig:Hc2}(a) and \ref{fig:Hc2}(b) for CuIr$_2$Te$_4$ and CuIr$_{1.95}$Ru$_{0.05}$Te$_4$, respectively.  
To determine the upper critical field at 0\,K, the $H_\mathrm{c2}(T)$ 
data were first analyzed by means of a Werthamer–Helfand–Hohenberg (WHH) model~\cite{Werthamer1966}. As shown by the solid lines in Fig.~\ref{fig:Hc2}, the WHH model can describe the experimental data 
reasonably well up to 0.075\,T. At higher magnetic fields, though,
this model fails to follow the data and underestimates $H_\mathrm{c2}$ values.
%
%
By contrast, $H_\mathrm{c2}(T)$ seem to show a linear temperature dependence.  
As indicated by the dash-dotted lines, a linear fit agrees remarkably well with the experimental data and provides $\mu_0H_\mathrm{c2}(0)$ = 0.212(1), and 0.241(1)\,T for CuIr$_2$Te$_4$ and CuIr$_{1.95}$Ru$_{0.05}$Te$_4$, respectively. 
Both $H_\mathrm{c2}(0)$ values are well below the weak-coupling Pauli-limit value (i.e., 1.86$T_c$ $\sim$ 5\,T), suggesting that the orbital pair-breaking effect is 
dominant in CuIr$_{2-x}$Ru$_x$Te$_4$ superconductors. 
A linear $H_\mathrm{c2}(T)$ over
a wide temperature range is uncommon.
Recently, e.g., it has been observed in infinite-layer nickelates La$_{1-x}$(Sr,Ca)$_x$NiO$_2$ and in noncentrosymmetric ThCo$_{1-x}$Ni$_x$C$_2$ superconductors~\cite{Sun2022,Chow2022,Grant2017}, 
the latter 
exhibiting line nodes in the superconducting gap~\cite{Bhattacharyya2019}. 
A linear $H_\mathrm{c2}(T)$ deviates significantly from
what is expected from a conventional BCS superconductor. As such, it strongly 
suggests an unconventional superconducting pairing in CuIr$_{2-x}$Ru$_x$Te$_4$.

In addition, in a single-band $s$-wave superconductor, the superfluid
density is almost indepedent of temperature for $T < 1/3T_c$. 
As a result, in general, $H_\mathrm{c2}$ saturates at low temperatures, as
clearly illustrated by the WHH model in Fig.~\ref{fig:Hc2}. 
Conversely, in a multiband superconductor, the superfluid density
in the non-dominant band (typically the one with the smaller gap)
increases with decreasing temperature (even below  $1/3T_c$), thus
leading to a continuously increasing $H_\mathrm{c2}$.
For example, typical multiband superconductors, such as MgB$_2$~\cite{Muller2001}
and Lu$_2$Fe$_3$Si$_5$~\cite{Nakajima2012}, exhibit a non-saturating
and almost linear $H_\mathrm{c2}(T)$. 
Based on this, we also analyzed $H_\mathrm{c2}(T)$ using a two-band
(TB) model~\cite{Gurevich2011}.
As indicated by dotted lines in Fig.~\ref{fig:Hc2}, the TB model, too,
shows a good agreement with the experimental data and yields comparable
$H_\mathrm{c2}(0)$ values. The multiband nature
of CuIr$_{2-x}$Ru$_x$Te$_4$ is
further confirmed by the temperature-dependent superfluid density and
electronic specific heat (see below), as well as by 
electronic band-structure calculations, where multiple bands
are identified to cross
the Fermi level~\cite{Yan2019}.

\subsection{Transverse-field- and zero-field $\mu$SR}
\begin{figure}[!thp]
	\centering
	\includegraphics[width=0.49\textwidth,angle= 0]{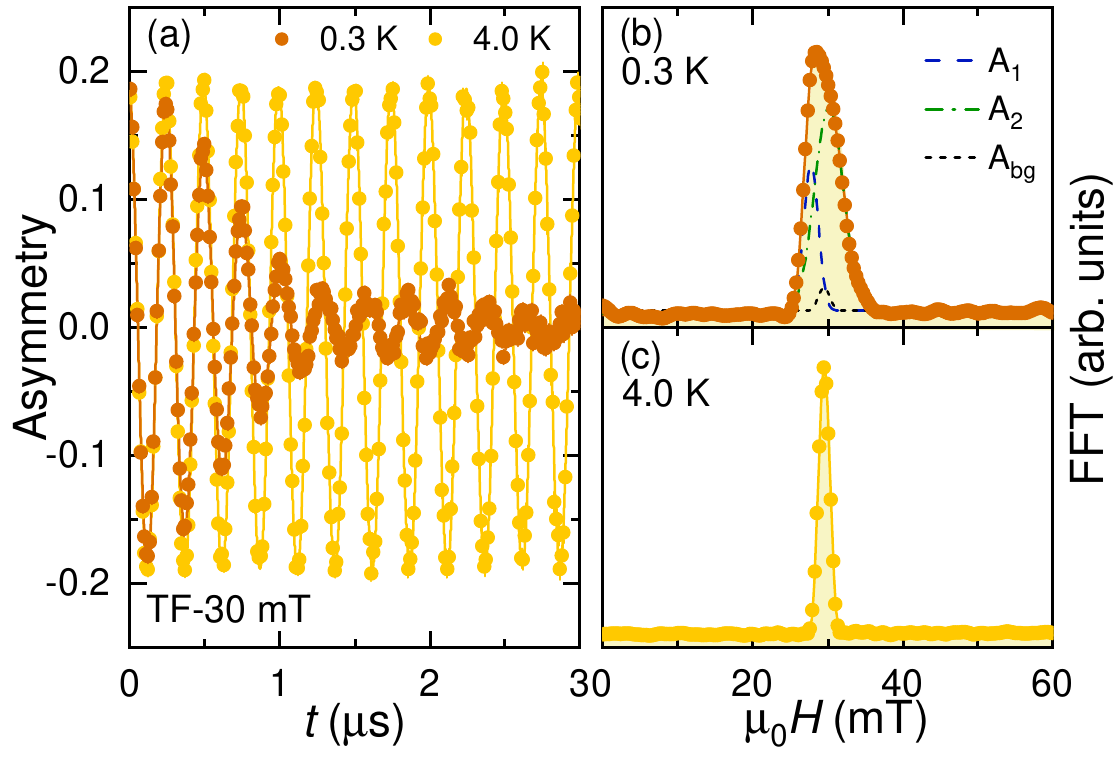}
	\caption{\label{fig:TF-muSR}(a) TF-$\mu$SR spectra collected in an
	applied field of 30\,mT in both the superconducting- and normal
	states of CuIr$_2$Te$_4$. 
	The respective real part of the Fourier transforms of $\mu$SR spectra are shown in (b) and (c) for 0.3\,K and 4.0\,K, respectively. 
    Solid lines are fits to Eq.~\eqref{eq:TF_muSR} using two oscillations. These are also shown separately as dashed- and dash-dotted lines in (b),
	together with a background contribution (dotted line).}
\end{figure}

Since our pure- and Ru-doped CuIr$_2$Te$_4$ samples share similar
features and there is no clear CDW transition in the pure case
(see below), most of our $\mu$SR measurements were performed on CuIr$_2$Te$_4$.
To investigate its superconducting pairing, we carried 
out systematic temperature-dependent TF-$\mu$SR measurements in applied magnetic fields of 30 and 80\,mT. 
After cooling in an applied field, the TF-$\mu$SR spectra were
collected upon heating. 
Representative TF-$\mu$SR spectra in the superconducting- and 
normal states of CuIr$_2$Te$_4$ are shown in Fig.~\ref{fig:TF-muSR}
for TF-30\,mT, while the TF-80\,mT $\mu$SR spectra are reported in
the Appendix~\ref{appendix}. In both cases, the normal-state spectra show essentially no damping, thus reflecting a uniform 
field distribution. Conversely, in the superconducting state (e.g., at 0.3\,K), the significantly enhanced damping reflects the 
inhomogeneous field distribution due to the development of a flux-line lattice (FLL)~\cite{Yaouanc2011,Amato1997,Blundell1999}. 
The broadening of the field distribution in the SC phase is clearly visible in 
Fig.~\ref{fig:TF-muSR}(b), where the fast-Fourier transform (FFT) spectra of the corresponding TF-30\,mT $\mu$SR data are shown.

To properly describe the field distribution, the TF-$\mu$SR spectra were modeled using~\cite{Maisuradze2009}:
\begin{equation}
	\label{eq:TF_muSR}
	A_\mathrm{TF}(t) = \sum\limits_{i=1}^n A_i \cos(\gamma_{\mu} B_i t + \phi) e^{- \sigma_i^2 t^2/2} +
	A_\mathrm{bg} \cos(\gamma_{\mu} B_\mathrm{bg} t + \phi).
\end{equation}
Here $A_{i}$ (97\%), $A_\mathrm{bg}$ (3\%) and $B_{i}$, $B_\mathrm{bg}$ 
are the initial asymmetries and local fields sensed by implanted muons in the 
sample and sample holder, $\gamma_{\mu}$/2$\pi$ = 135.53\,MHz/T 
is the muon gyromagnetic ratio, $\phi$ is a shared initial phase, and $\sigma_{i}$ 
is the Gaussian relaxation rate of the $i$th component. 
Here, we find that, 
while two oscillations (i.e., $n = 2$) are required to properly describe the TF-30\,mT $\mu$SR spectra, a single oscillation is sufficient for the 80-mT spectra
(see details in Appendix~\ref{appendix}). 
%
%
%
\begin{figure}[!thp]
	\centering
	\includegraphics[width=0.44\textwidth,angle= 0]{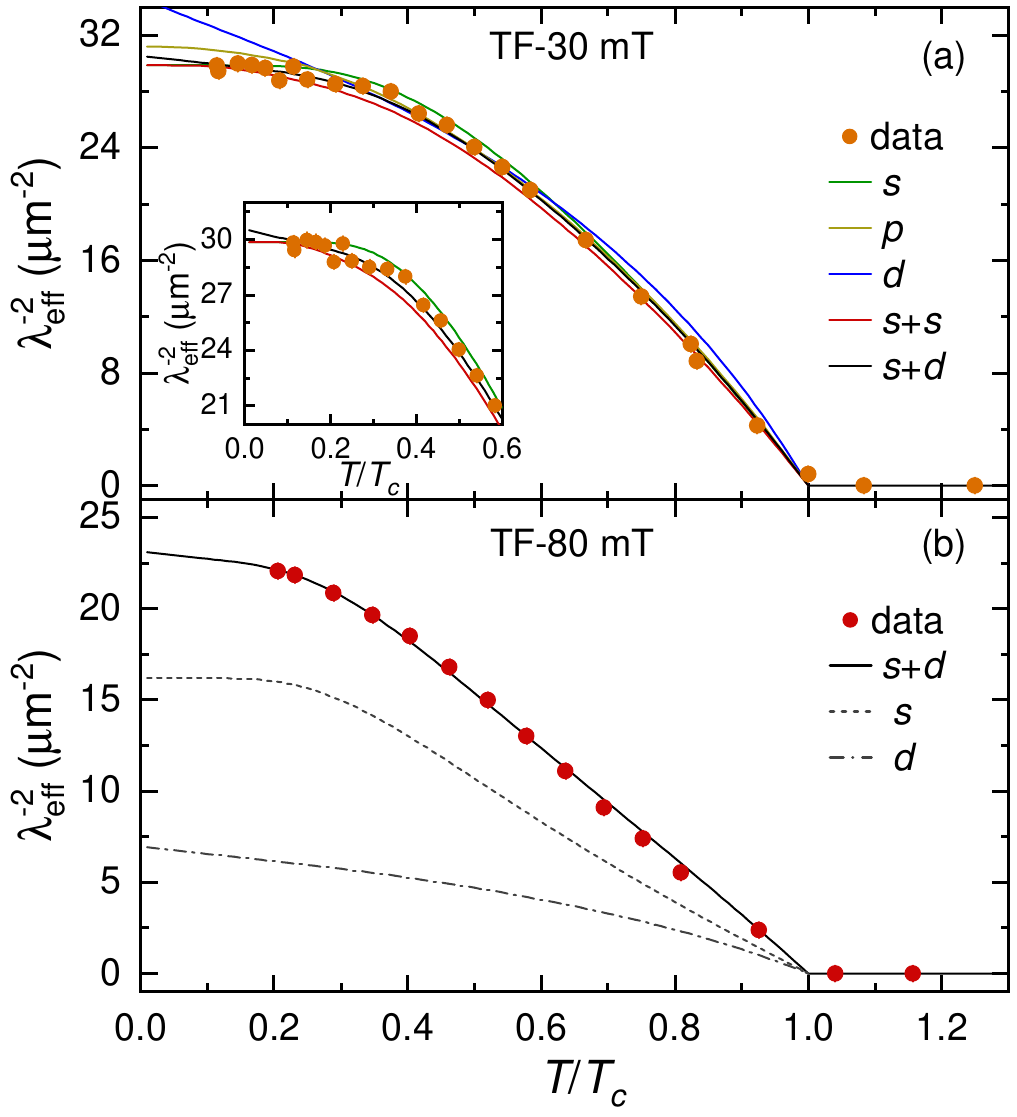}
	\caption{\label{fig:superfluid}Superfluid density vs temperature, as determined from
		TF-$\mu$SR measurements in a magnetic field of 30\,mT (a) and 80\,mT (b). The inset in (a) shows the enlarged plot below $T/T_c$ = 0.6. 
		The different lines represent fits to various models, including single-gap $s$-, $p$-, and $d$-wave, and two-gap ($s+s$)- and ($s+d$)-wave (see text for details). 
	    In the latter case, the $s$- (dashed line) and $d$-components
		(dash-dotted line) are also shown separately in panel (b) for 
		the TF-80\,mT case. All fit parameters are listed in Table~\ref{tab:table1}.}
\end{figure}
%
%
In the 30-mT case, the dashed-, dash-dotted-, and dotted lines in Fig.~\ref{fig:TF-muSR}(b) represent the two components at 0.3\,K ($A_1$ and $A_2$) and the background signal ($A_\mathrm{bg}$), respectively.
A similar behavior has been found in other superconductors, e.g., Mo$_3$P or ReBe$_{22}$~\cite{Shang2019b,Shang2019c}, where the $\mu$SR spectra collected at
higher magnetic fields exhibit a more symmetric field distribution.
For TF-30mT $\mu$SR, the effective 
Gaussian relaxation rate $\sigma_\mathrm{eff}$ can be calculated
from $\sigma_\mathrm{eff}^2/\gamma_\mu^2 = \sum_{i=1}^2 A_i [\sigma_i^2/\gamma_{\mu}^2 - \left(B_i - \langle B \rangle\right)^2]/A_\mathrm{tot}$~\cite{Maisuradze2009}, where  $\langle B \rangle = (A_1\,B_1 + A_2\,B_2)/A_\mathrm{tot}$ and $A_\mathrm{tot} = A_1 + A_2$.
Considering the constant nuclear relaxation rate $\sigma_\mathrm{n}$ 
in the narrow temperature range investigated here, confirmed also 
by ZF-$\mu$SR measurements (see below), 
the superconducting Gaussian relaxation rate can be extracted using 
$\sigma_\mathrm{sc} = \sqrt{\sigma_\mathrm{eff}^{2} - \sigma^{2}_\mathrm{n}}$. 
Then, the superconducting gap value and its symmetry can be investigated by measuring the temperature-dependent $\sigma_\mathrm{sc}$, which is directly related to the magnetic penetration depth and thus the superfluid density. 
Since the upper critical field of CuIr$_2$Te$_4$  ($\sim$ 0.2\,T) is not significantly large compared to the applied TF fields (30 and 80\,mT), the effective penetration depth $\lambda_\mathrm{eff}$
had to be calculated from $\sigma_\mathrm{sc}$ by considering the overlap
of the vortex cores. 
Consequently, in our case, $\lambda_\mathrm{eff}$ was calculated by means
of $\sigma_\mathrm{sc} = 0.172 \frac{\gamma_{\mu} \Phi_0}{2\pi}(1-h)[1+1.21(1-\sqrt{h})^3]\lambda_\mathrm{eff}^{-2}$~\cite{Barford1988,Brandt2003}, where $h = H_\mathrm{appl}/H_\mathrm{c2}$,
with $H_\mathrm{appl}$ 
the applied magnetic field.

%
%
\begin{figure}[!htp]
	\centering
	\includegraphics[width=0.45\textwidth,angle= 0]{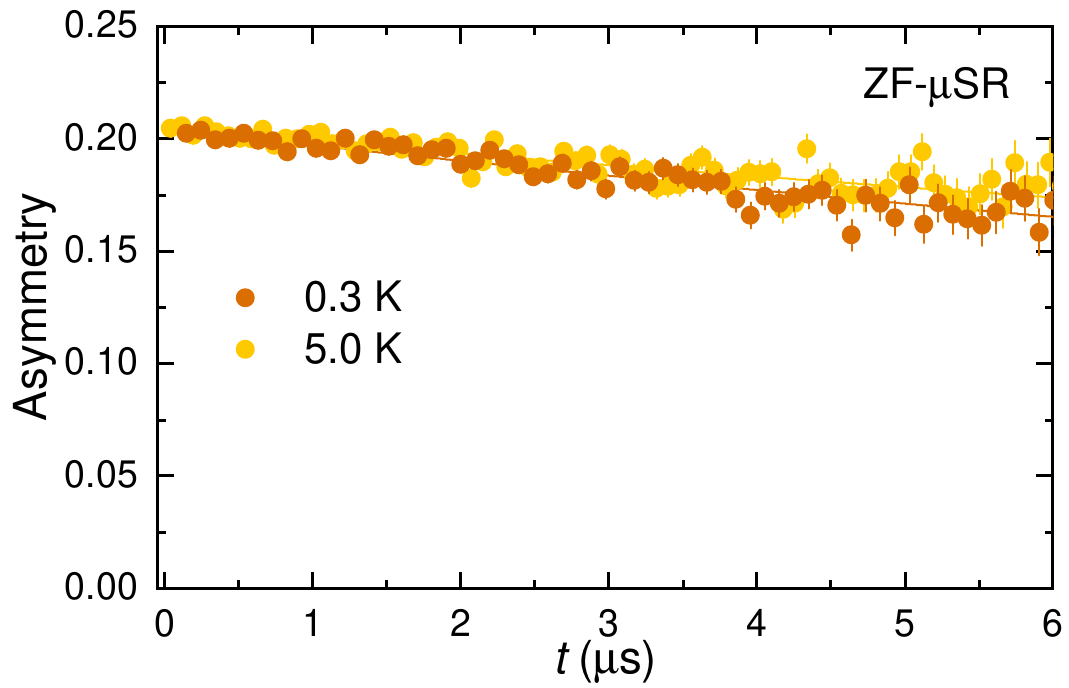}
	\caption{\label{fig:ZF-muSR}ZF-$\mu$SR spectra collected in the superconducting- (0.3\,K) and the normal (5\,K) states of CuIr$_2$Te$_4$.
		The practically overlapping datasets indicate the absence of 
		TRS breaking, whose occurrence would have resulted in a stronger 
		decay in the 0.3-K case.}
\end{figure}
%
 
The inverse square of the magnetic penetration depth [proportional to the superfluid density, i.e., $\lambda_\mathrm{eff}^{-2}(T) \propto \rho_\mathrm{sc}(T)$] 
vs. the reduced temperature $T/T_c$ is shown in Fig.~\ref{fig:superfluid}(a) and \ref{fig:superfluid}(b) for TF-30\,mT and TF-80\,mT, respectively.
In both cases, the superfluid density remains weakly temperature dependent down to the lowest temperature, i.e., below 1/3$T_c$.  
Such behavior indicates the presence of low-energy excitations and, hence, of nodes in the superconducting gap.  
To get further insight into the pairing symmetry, the superfluid density $\rho_\mathrm{sc}(T)$ was analyzed using different models, generally described by: 
\begin{equation}
	\label{eq:rhos}
	\rho_\mathrm{sc}(T) = 1 + 2\, \Bigg{\langle} \int^{\infty}_{\Delta_\mathrm{k}} \frac{E}{\sqrt{E^2-\Delta_\mathrm{k}^2}} \frac{\partial f}{\partial E} \mathrm{d}E \Bigg{\rangle}_\mathrm{FS}. 
\end{equation}
Here, $f = (1+e^{E/k_\mathrm{B}T})^{-1}$ is the Fermi function and
$\langle \rangle_\mathrm{FS}$ represents an average over the Fermi surface~\cite{Tinkham1996}. 
$\Delta_\mathrm{k}(T) = \Delta(T) g_\mathrm{k}$ is the product of
$\Delta(T)$, the temperature-dependent gap, and $g_\mathrm{k}$, the 
angular dependence of the gap (see details in Table~\ref{tab:table1}). 
The temperature dependence of the gap is assumed to follow $\Delta(T) = \Delta_0 \mathrm{tanh} \{1.82[1.018(T_\mathrm{c}/T-1)]^{0.51} \}$~\cite{Tinkham1996,Carrington2003}, where $\Delta_0$ is the gap value at 0\,K.

Five different models, including single-gap $s$-, $p$-, and $d$-wave,
and two-gap ($s+s$)- and ($s+d$)-wave, were used to analyze the 
$\lambda_\mathrm{eff}^{-2}$$(T)$ data. The derived fitting parameters 
are listed in Table~\ref{tab:table1}. 
As can be clearly seen in Fig.~\ref{fig:superfluid}(a), the weak
temperature dependence of the superfluid density at low-$T$ rules out
a line-node $d$-wave model (see blue line). 
In case of an $s$- or $p$-wave model, we also find a poor agreement
with the data below $T/T_c$ $\sim$ 0.3 (see green and yellow lines). 
For the two-gap scenario,
we consider here the so-called $\alpha$-model. 
In this case, the superfluid density can be described by $\rho_\mathrm{sc}(T) = w \rho_\mathrm{sc}^{\Delta^\mathrm{f}}(T) + (1-w) \rho_\mathrm{sc}^{\Delta^\mathrm{s}}(T)$, 
where $\rho_\mathrm{sc}^{\Delta^\mathrm{f}}$ and 
$\rho_\mathrm{sc}^{\Delta^\mathrm{s}}$ are the superfluid densities related to 
the first ($\Delta^\mathrm{f}$) and second ($\Delta^\mathrm{s}$) gaps, and $w$ is a 
relative weight. For each gap, $\rho_\mathrm{sc}(T)$ is given by 
Eq.~\eqref{eq:rhos}.
The superfluid density is best fitted by a two-gap ($s+d$)-wave model (see black line), while the ($s+s$)-wave model (see red line) shows a clear deviation from the low-$T$ data [see enlarged plot in the inset of Fig.~\ref{fig:superfluid}(a)]. This is also reflected in the smallest $\chi_\mathrm{r}^2$ value for the ($s+d$)-wave model (see details in Table~\ref{tab:table1}).
It is noted that, though in general the ($s+p$)-wave model also can describe the data reasonably well, it is inconsistent with the preserved time-reversal symmetry (TRS) in the superconducting state of CuIr$_2$Te$_4$ (see below). 
This is different from the case of CaPtAs superconductor, where the ($s+p$)-wave model was proposed to account for both the gap nodes and broken TRS in the superconducting state~\cite{Shang2020}.   
%
%
\begin{figure}[!thp]
	\centering
	\includegraphics[width=0.45\textwidth,angle=0]{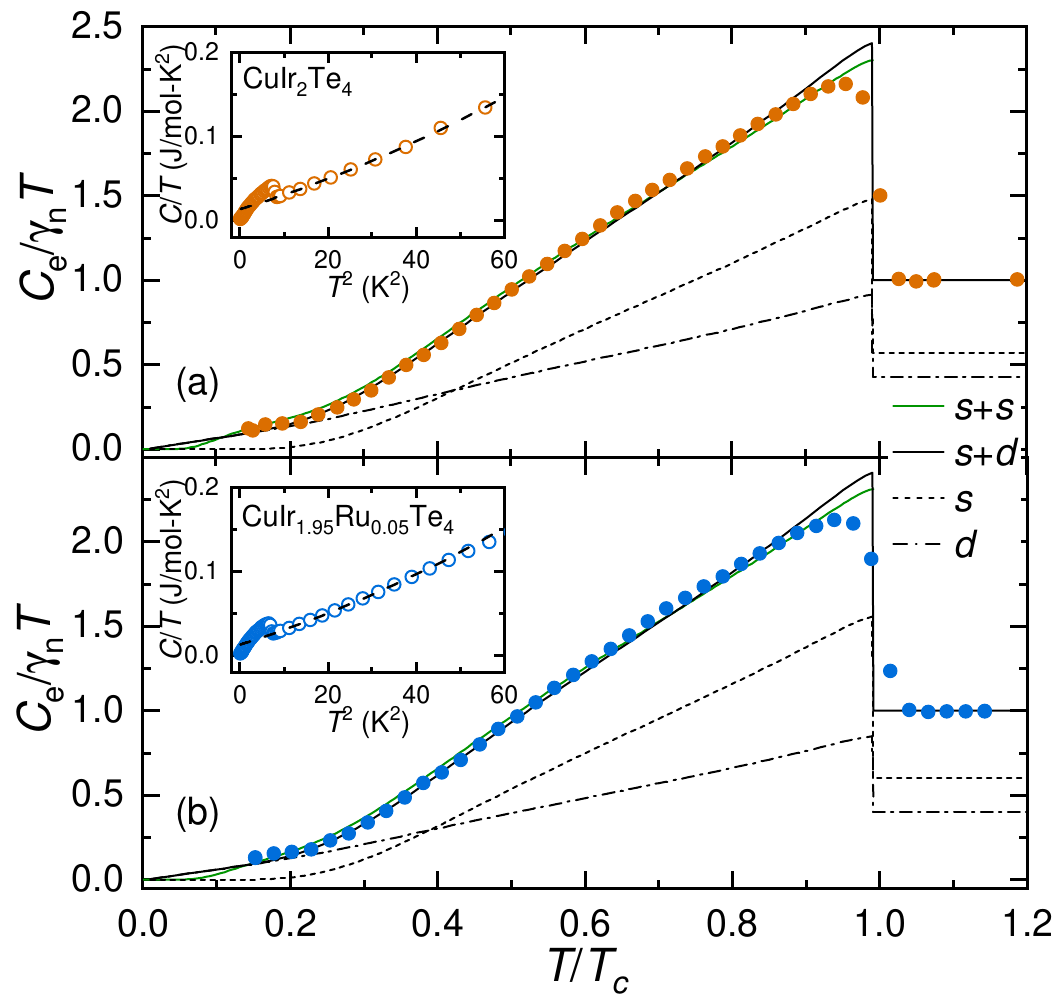}
	\vspace{-2ex}%
	\caption{\label{fig:specific_heat}Normalized electronic specific heat $C_\mathrm{e}$/$\gamma_\mathrm{n}T$ as a function of reduced temperature $T/T_c$
		for CuIr$_2$Te$_4$ (a) and CuIr$_{1.95}$Ru$_{0.05}$Te$_4$ (b).
		The insets show the measured specific heat $C/T$ versus $T^2$. The dashed lines in the insets are fits to $C/T = \gamma_\mathrm{n} + \beta T^2 + \delta T^4$ for $T > T_c$. 
		The solid green and black lines represent the
		electronic specific heat calculated by considering a two-gap ($s+s$)- and ($s+d$)-wave model, respectively. The dashed- and dash-dotted lines show 
		the individual contributions from the $s$- and $d$-type	SC gaps
		for the ($s+d$)-wave model.
		The fitting parameters are listed in Table~\ref{tab:table1}.}
\end{figure}
%
%
In the TF-80\,mT $\mu$SR case [see Fig.~\ref{fig:superfluid}(b)], the increased magnetic field suppresses the $s$-type gap from 1.75 to 1.3\,$k_\mathrm{B}T_c$, while the $d$-type gap and its weight remain the same (see weights reported 
in Table~\ref{tab:table1}). The separate $s$- and $d$-components of the superfluid density are shown by dotted-
and dash-dotted lines in Fig.~\ref{fig:superfluid}(b). 
The suppression of the $s$-type gap at 80\,mT makes
the nodal features more evident in the superfluid density, nevertheless,
further low-$T$ measurements (below 0.3\,K) are crucial.

To search for a possible breaking of the time-reversal symmetry (TRS)
in the superconducting state of CuIr$_2$Te$_4$, zero-field (ZF-) $\mu$SR
measurements were performed in its normal- and superconducting states.  
As shown in Fig.~\ref{fig:ZF-muSR}, neither coherent oscillations nor fast decays could be identified in the spectra collected below (0.3\,K) and above $T_c$ (5\,K),
thus excluding any type of magnetic order or fluctuations.
In case of nonmagnetic materials,
in the absence of applied fields, the depolarization of muon spins
is mainly determined by the randomly oriented nuclear magnetic moments. 
In CuIr$_2$Te$_4$, the depolarization shown in Fig.~\ref{fig:ZF-muSR}
is more consistent with a Lorentzian decay. This suggests that the internal fields sensed by the implanted muons arise from the diluted (and tiny) nuclear
moments present in CuIr$_2$Te$_4$. Thus, the solid lines in Fig.~\ref{fig:ZF-muSR} are fits to a Lorentzian Kubo-Toyabe relaxation function
$A(t) = A_\mathrm{s}[\frac{1}{3} + \frac{2}{3}(1 - \Lambda_\mathrm{ZF} t) \mathrm{e}^{- \Lambda_\mathrm{ZF} t}] + A_\mathrm{bg}$. 
Here, $A_\mathrm{s}$ and $A_\mathrm{bg}$ are the same as in the 
TF-$\mu$SR case [see Eq.~\eqref{eq:TF_muSR}], while
$\Lambda_\mathrm{ZF}$ represents the ZF Lorentzian relaxation rate. 
The derived relaxation rates in the normal- and the superconducting state are almost identical, i.e., $\Lambda_\mathrm{ZF}$  = 0.0263(16)\,$\mu$s$^{-1}$ at 0.3\,K, and $\Lambda_\mathrm{ZF}$ = 0.0267(14)\,$\mu$s$^{-1}$ at 5\,K, 
as also reflected in the overlapping datasets. The lack of additional $\mu$SR relaxation below $T_c$ excludes a possible TRS
breaking in the superconducting state of CuIr$_2$Te$_4$.
As a consequence, by taking into account the preserved TRS in
CuIr$_2$Te$_4$, we have to exclude the ($s+p$)-wave model, since
only the ($s+d$)-wave model is compatible with the experiment.

\begin{table*}[tbp]
	\centering
	\caption{Summary of the superfluid-density- and electronic
	specific-heat data analysis using different models for CuIr$_2$Te$_4$. 
	In the gap function $g_\mathrm{k}$, $\theta$ and $\phi$ are the polar and azimuthal angles in $k$-space.
	The SC gap values are in meV units, while the zero-temperature
	magnetic penetration depths $\lambda_0$ are in nm.
	In the last column, the reduced least-square deviations $\chi_\mathrm{r}^2$ are reported for the 
	TF-$\mu$SR (30\,mT) and $C_\mathrm{e}/T$ data, respectively. The weights listed in table refer to the first $s$-wave component. \label{tab:table1}} 
	\begin{ruledtabular}
		\begin{tabular}{lcccccc}
			\textrm{Model}&
			\textrm{$g_\mathrm{k}$ function}&
			\textrm{Gap type}&
		    \textrm{$\lambda_0$}&
			\textrm{$\Delta_0$(TF-$\mu$SR)}&
			\textrm{$\Delta_0$($C_\mathrm{e}$/$T$)}&
			\textrm{$\chi_\mathrm{r}^2$} ($\mu$SR / $C_\mathrm{e}/T$)\\
			\hline
			\vspace{1pt}
			$s$-wave       & 1                 & nodeless          & 183        & 0.36                 & 0.43                              & 6.4 / >30 \\  
			\vspace{1pt}    
			$p$-wave       & sin$\theta$       & point-node        & 179        & 0.46                 & 0.51                               & 7.8 / 14.4\\
			\vspace{1pt}   
	    	$d$-wave       & cos2$\phi$        & line-node         & 170        & 0.54                 & 0.57                              & 23.1 / >30 \\ 
			\vspace{1pt}
			 $s+s$ (weighted)     & 1, 1               & nodeless        & 183       & 0.13/0.37(0.1)    & 0.12/0.44(0.15)                         & 6.3 / 9.5 \\
			\vspace{1pt} 
			$s+d$ (weighted)     & 1, cos2$\phi$     & line-node         & 181        & 0.36/0.52(0.70)    & 0.45/0.54(0.57)                      & 4.7 / 5.1 \\
		\end{tabular} 
	\end{ruledtabular}
\end{table*}

\subsection{Electronic specific heat}

To further validate the superconducting pairing of CuIr$_{2-x}$Ru$_x$Te$_4$,
its zero-field electronic specific heat $C_e/T$ was analyzed using the aforementioned
models. 
To subtract the phonon contribution from the specific heat, 
the normal-state specific heat was fitted to the expression $C/T = \gamma_\mathrm{n} + \beta T^2 + \delta T^4$, 
with $\gamma_\mathrm{n}$ the normal-state electronic specific-heat 
coefficient, $\beta$ and $\delta$ the phonon specific-heat coefficients
(see dashed lines in the insets of Fig~\ref{fig:specific_heat}). 
From this, we derive $\gamma_\mathrm{n}$ = 13.1(7)\,mJ/mol-K$^2$, $\beta$ = 1.63(6)\,mJ/mol-K$^4$, and $\delta$ = 0.010(1)\,mJ/mol-K$^6$ 
for CuIr$_2$Te$_4$, while for CuIr$_{1.95}$Ru$_{0.05}$Te$_4$, $\gamma_\mathrm{n}$ = 12.6(1)\,mJ/mol-K$^2$, $\beta$ = 1.66(1)\,mJ/mol-K$^4$, and $\delta$ = 0.011(1)\,mJ/mol-K$^6$.
After subtracting the phonon contribution ($\beta$$T^2$ + $\delta$$T^4$)
from the raw data, the electronic specific heat divided by $\gamma_\mathrm{n}$, i.e., $C_e/\gamma_\mathrm{n}T$, is obtained.
This is shown in Fig.~\ref{fig:specific_heat} vs.\ the reduced temperature $T/T_c$ for both CuIr$_2$Te$_4$ and
CuIr$_{1.95}$Ru$_{0.05}$Te$_4$.

The contribution of the superconducting phase to entropy can
be calculated following the BCS expression~\cite{Tinkham1996}:
\begin{equation}
	\label{eq:entropy}
	S(T) = -\frac{6\gamma_\mathrm{n}}{\pi^2 k_\mathrm{B}} \int^{\infty}_0 [f\mathrm{ln}f+(1-f)\mathrm{ln}(1-f)]\,\mathrm{d}\epsilon,
\end{equation}
where $f$ is the same as in Eq.~\eqref{eq:rhos}.
Then, the temperature-dependent electronic specific heat in the superconducting 
state can be calculated from $C_\mathrm{e} =T \frac{dS}{dT}$. 
In case of a multiple-gap model, the electronic specific heat can 
be modeled by $C_\mathrm{e}(T)/T = wC_\mathrm{e}^{\Delta^\mathrm{f}}(T)/T + (1-w)C_\mathrm{e}^{\Delta^\mathrm{s}}(T)/T$~\cite{Bouquet2001}. 
Here, each term represents the contribution to the specific heat
of the individual gaps, with $w$, $\Delta^\mathrm{f}$, and $\Delta^\mathrm{s}$ being the
same parameters as for the superfluid-density fits.
To analyze the electronic specific heat, we employ the same models 
used to fit the superfluid density. The fit parameters obtained in both
cases are listed in Table~\ref{tab:table1}. Also for the specific heat, the single-gap $s$-, $p$-, and $d$-wave models deviate significantly from the data, here reflected in larger $\chi_\mathrm{r}^2$ values.
Conversely, the multigap models exhibit a much better agreement with the
experimental data across the full temperature range, with the ($s+d$)-wave model (solid-black lines) showing the smallest deviation (i.e., smallest $\chi_\mathrm{r}^2$).
While the ($s+s$)-wave model (solid-green lines) reproduces the data for $T/T_c \gtrsim 0.6$, 
it deviates from them at low temperatures, hence yielding a slightly larger $\chi_\mathrm{r}^2$ than the two-gap ($s+d$)-wave model (see Table~\ref{tab:table1}).
In summary, both the temperature-dependent superfluid density and 
electronic specific heat are well described by a two-gap ($s+d$)-wave model, 
hence providing strong evidence about nodal superconductivity
in CuIr$_{1-x}$Ru$_{x}$Te$_4$. Incidentally, the multigap features
are also reflected in the temperature-dependent upper critical
field of CuIr$_{1-x}$Ru$_{x}$Te$_4$ (see details in Fig.~\ref{fig:Hc2}).

%
%
\begin{figure}[!thp]
	\centering
	\includegraphics[width=0.48\textwidth,angle=0]{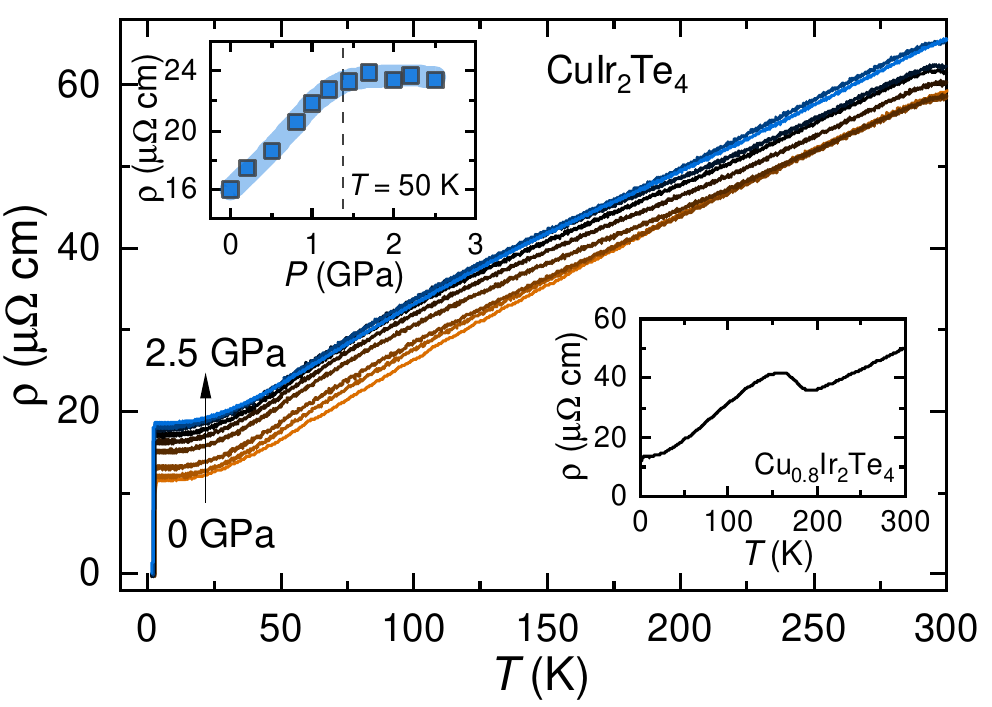}
	\vspace{-2ex}%
	\caption{\label{fig:rho_p} Temperature-dependent electrical resistivity
	measured under various external pressures up to 2.5\,GPa for CuIr$_2$Te$_4$.
	The upper inset summarizes the 50-K electrical-resistivity
	vs.\ pressure. The lower inset shows the
	electrical resistivity for Cu$_{0.8}$Ir$_2$Te$_4$, where the
	hump at $\sim 190$\,K indicates the CDW transtion.}
\end{figure}
%
%

\subsection{Pressure effects}

We also investigated the pressure effects on the normal- and superconducting states of CuIr$_2$Te$_4$. As shown in Fig.~\ref{fig:rho_p}, the temperature-dependent electrical resistivity was measured 
at various external pressures up to 2.5\,GPa. At ambient pressure, 
$\rho(T)$ shows the typical behavior of metallic compounds,
with no peculiar features related to possible phase transitions in the normal state.
Previous studies report a dramatic jump in $\rho(T)$ near 250\,K, attributed to the CDW transition~\cite{Yan2019,Dong2019,Nagata1999}.
In our case, the absence of a CDW anomaly might be due to a slightly different Cu-content.
Indeed, our preliminary electrical-resistivity measurements on
Cu$_{0.8}$Ir$_2$Te$_4$ reveal a clear CDW transition at $\sim 190$\,K
(see bottom inset in Fig.~\ref{fig:rho_p}).
As the pressure increases to $\sim 1.2$\,GPa,
$\rho(T)$ exhibits a broad hump around 100\,K, most likely related
to the CDW transition. As the pressure increases further, the hump
shifts to higher temperatures. The top inset in Fig.~\ref{fig:rho_p}
summarizes the dependence of the 50-K electrical resistivity on external pressure. 
First, the electrical resistivity increases with pressure. Then, above 1.2\,GPa, where the resistive hump becomes more evident, it starts to saturate. Such behavior is consistent with 
previous chemical-pressure studies on CuIr$_2$Te$_4$, indicating that both S/Te and Se/Te substitutions favor the CDW order~\cite{Boubeche2021,Boubeche2022}. 

%
%
\begin{figure}[!thp]
	\centering
	\includegraphics[width=0.46\textwidth,angle=0]{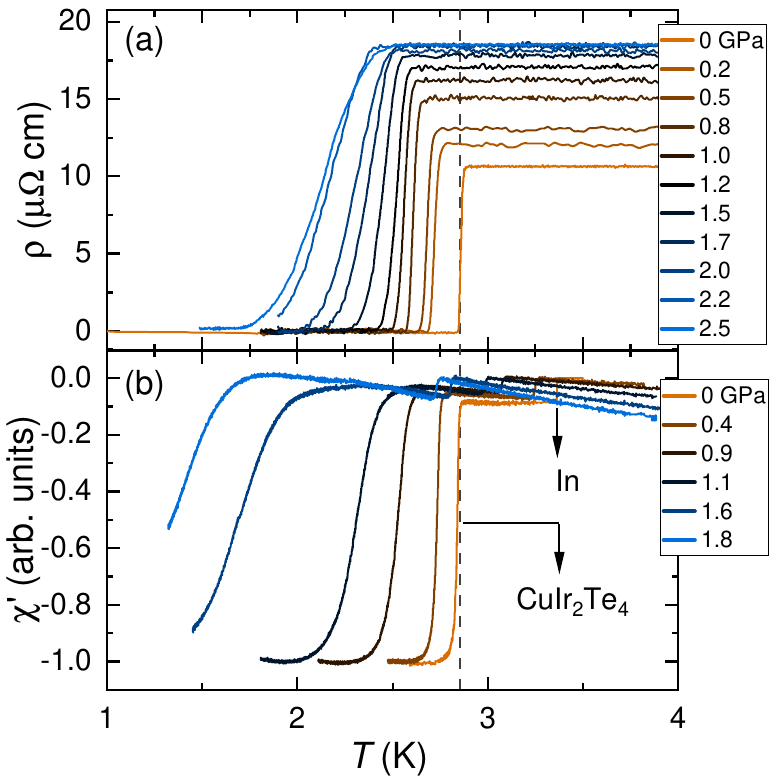}
	\vspace{-2ex}%
	\caption{\label{fig:chi_p}Low-$T$ electrical resistivity $\rho(T)$ (a)
	and real part of ac susceptibility $\chi'(T)$ (b), collected under 
	applied pressure up to 2.5\,GPa.
	$T_c$ was defined as the onset of zero resistivity in the $\rho(T)$
	curves, or as the onset of superconducting transition in the $\chi'(T)$ curves
	(both marked by a dashed line for the ambient-pressure case).
	The SC transition of indium is used to determine the applied pressure 
	during the $\chi'(T)$ measurements.}
\end{figure}
%
Figure~\ref{fig:chi_p} shows the low-$T$ (below 4\,K) electrical
resistivity and ac susceptibility collected at
various applied
pressures. As the pressure increases, the superconducting transition,
in both $\rho(T)$ and $\chi'(T)$, becomes broader and $T_c$ is
progressively suppressed to lower temperatures. Similar to the
ambient-pressure case (see details in Fig.~\ref{fig:Hc2_raw}), the $T_c$ is again defined as the onset
of zero resistivity. This coincides with the onset of superconducting
transition in $\chi'(T)$ and is indicated by dashed lines in the two panels of Fig.~\ref{fig:chi_p}.  
These highly consistent $T_c$ values versus the applied pressure are summarized in Fig.~\ref{fig:pressure_phase}(a).
The $T_c(P)$ exhibits a decreasing nonlinear trend, 
starting at 2.85\,K at ambient pressure to reach 1.72\,K at 2.5\,GPa.
As shown in Fig.~\ref{fig:pressure_phase}(b), we measured also the electrical resistivity 
at 2.2\,GPa under increasingly higher magnetic fields, up to 0.15\,T. 
Interestingly, the $H_\mathrm{c2}(T)$ under applied pressure is significantly different from the ambient-pressure
case. $H_\mathrm{c2}(T)$ at 2.2\,GPa is well described by the WHH model [see solid line in Fig.~\ref{fig:pressure_phase}(c)],
more consistent with a single-band $s$-wave superconductor.
Conversely, at ambient pressure, the linear dependence of
$H_\mathrm{c2}(T)$ is attributed to multiple gaps and unconventional pairing.
This suggests that pressure most likely suppresses the
nodal component of superconductivity, thus changing its character 
from partially nodal towards fully gapped.

\subsection{Discussion}
According to the temperature-dependent superfluid density and zero-field
electronic specific-heat data, CuIr$_2$Te$_4$ exhibits a multigap SC, 
best described by an ($s+d$)-wave model. 
%
\begin{figure}[!thp]
	\centering
	\includegraphics[width=0.5\textwidth,angle=0]{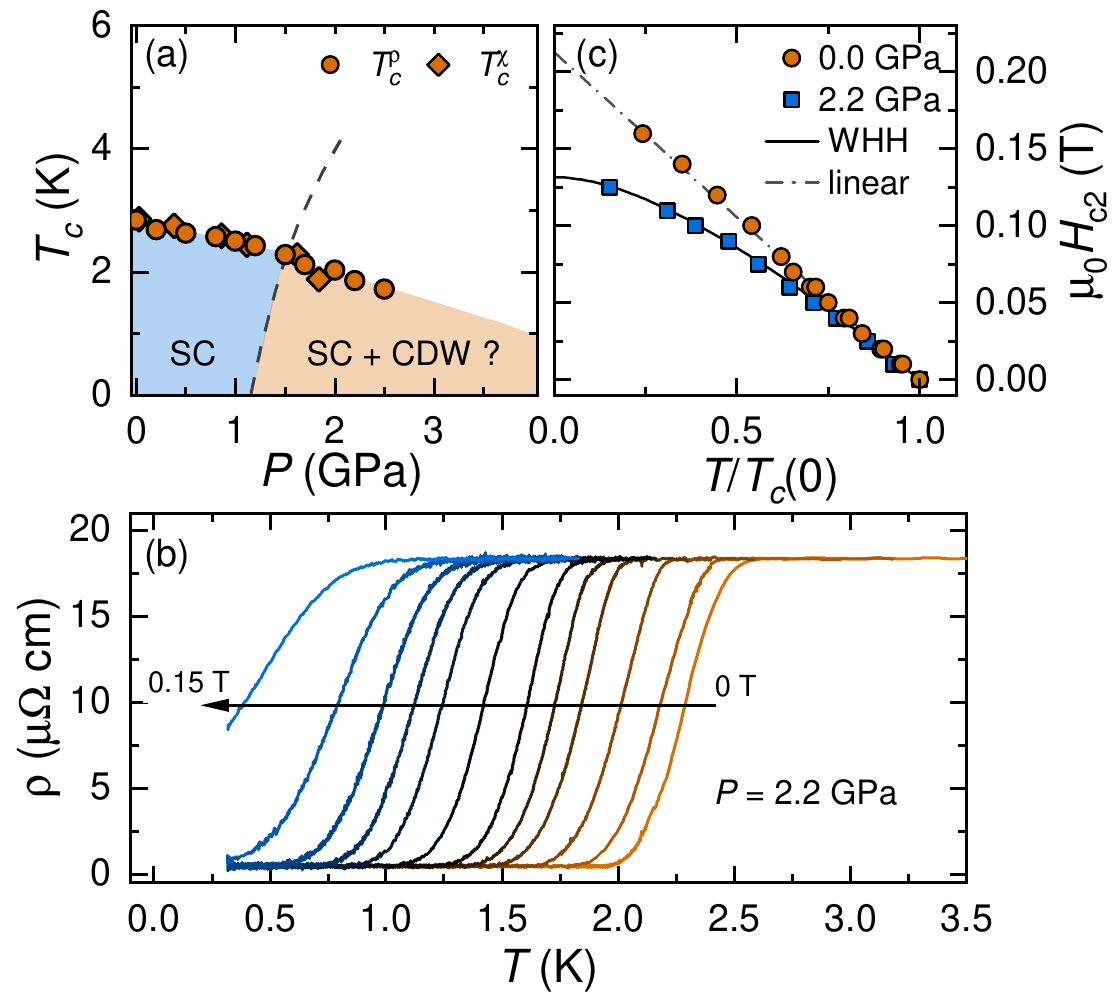}
	\vspace{-2ex}%
	\caption{\label{fig:pressure_phase}(a) Superconducting transition
	temperature $T_c$ vs.\ external pressure. $T_c$ was determined
	from the measurements shown in Fig.~\ref{fig:chi_p}. 
	(b) Temperature-dependent electrical resistivity at 2.2\,GPa
	measured under various magnetic fields, up to 0.15\,T. 
	 (c) Upper critical field $H_\mathrm{c2}$ vs.\ the reduced transition
	temperature $T_c/T_c(0)$ for CuIr$_2$Te$_4$ at ambient pressure
	and at 2.2\,GPa. The solid line represents a fit to the WHH model, while the
	dash-dotted line indicates a linear temperature dependence.}
\end{figure}
%
In both cases, the presence of gap nodes and thus, of
low-energy excitations, is reflected in a weak (i.e., non constant) temperature
dependence for $T < 1/3T_c$ (see details in Fig.~\ref{fig:superfluid} and Fig.~\ref{fig:specific_heat}).  
The multigap nature of SC is further confirmed by the temperature-dependent upper critical field (see Fig.~\ref{fig:Hc2}).
In both CuIr$_2$Te$_4$ and CuIr$_{0.95}$Ru$_{0.05}$Te$_4$, the two-band model is 
clearly superior to the WHH model in the low-$T$ and/or high-field region. In the CuIr$_2$Te$_4$ case, also the electronic band-structure calculations support a multigap SC, since they indicate that multiple bands cross the Fermi level~\cite{Yan2019}.
In our case, a linear $H_\mathrm{c2}(T)$ over a wide temperature
range departs significantly from the $H_\mathrm{c2}(T)$ of most BCS superconductors
and is likely attributed to the presence of nodes in the superconducting gap. 
As recently shown in ThCo$_{1-x}$Ni$_x$C$_2$ superconductors, a
linear $H_\mathrm{c2}(T)$ was proposed to be closely related to a $d$-wave pairing~\cite{Grant2017,Bhattacharyya2019}. 
Which electronic bands account for the $s$-wave and $d$-wave pairing in CuIr$_2$Te$_4$ is not yet known and requires further theoretical investigation.

$H_\mathrm{c2}(T)$ measured under applied pressure differs
significantly from that measured at ambient pressure. For instance,
at 2.2\,GPa, $H_\mathrm{c2}(T)$ follows very well the WHH model
[see details in Fig.~\ref{fig:pressure_phase}(c)], more consistent
with a single-band $s$-wave superconductor.
As proposed in Fig.~\ref{fig:pressure_phase}(a), in the low-pressure region, CuIr$_2$Te$_4$ shows a pure
superconducting phase below $T_c$. 
As the pressure increases towards 1.2\,GPa, a CDW order starts
to develop and, at higher applied pressures, the SC phase might
coexist with the CDW phase, hence, becoming more conventional.
The dashed line in Fig.~\ref{fig:pressure_phase}(a) separates these
qualitatively different SC phases. Eventually, at even higher
pressures (around 4.5\,GPa) the CDW phase becomes dominant and entirely
suppresses the SC phase (to be confirmed experimentally). 
A similar phase diagram is shown by the isostructural
Ir$_{0.95}$Pt$_{0.05}$Te$_2$ compounds, where again the external
pressure suppresses the superconductivity and gives rise to a
CDW order~\cite{Ivashko2017}. In the low-pressure region, the
unconventional ($s+d$)-wave pairing of
CuIr$_2$Te$_4$ might reflect the ubiquitous charge fluctuations near
the CDW quantum critical point. A further increase in pressure quenches these fluctuations and
makes the $s$-wave pairing more favorable.
By contrast, in the (Ca$_{1-x}$Sr$_x$)$_3$Ir$_4$Sn$_{13}$ family,
despite a similar phase diagram to CuIr$_2$Te$_4$~\cite{Klintberg2012},
the superconducting order parameter of both the pure-SC and
the SC+CDW phase maintains the same $s$-wave character~\cite{Biswas2015}.
Consequently, CuIr$_2$Te$_4$ might represent a rare case, where the SC and SC+CDW phases show different superconducting pairings.  
To prove such scenario, it would be interesting to investigate
the evolution of the superconducting pairing in CuIr$_2$Te$_4$
under applied pressure, e.g., via TF-$\mu$SR measurements under pressure. 
Alternatively, TF-$\mu$SR measurements on CuIr$_2$Te$_{4-x}$(Se,S)$_x$, where multiple CDW phases are known to coexist with SC~\cite{Boubeche2021,Boubeche2022},
would also provide important hints on the superconducting pairing.

\section{\label{ssec:Sum}Conclusion}\enlargethispage{8pt}

In summary, we investigated the normal- and the superconducting properties
of CuIr$_{2-x}$Ru$_x$Te$_4$ ($x$ = 0, 0.05) by means of electrical resistivity-, magnetization-, heat ca\-pac\-i\-ty-, and $\mu$SR measurements.
CuIr$_2$Te$_4$ and CuIr$_{1.95}$Ru$_{0.05}$Te$_4$ exhibit
bulk superconductivity with $T_c$ = 2.85 and 2.7\,K, respectively. 
Both the temperature-dependent superfluid density and the electronic
specific heat are best described by a two-gap 
model [here, ($s+d$)-wave], comprising a nodeless gap and a gap with nodes,
rather than by single-band models. 
The multigap SC in CuIr$_{2-x}$Ru$_x$Te$_4$ is further supported by the temperature dependence of the
upper critical field $H_{c2}(T)$. 
The application of external pressure promotes the formation
of CDW order and shifts CuIr$_2$Te$_4$ towards a conventional $s$-wave SC behavior. The unconventional
superconducting pairing in CuIr$_{2-x}$Ru$_x$Te$_4$ seems closely 
related to the charge fluctuations occurring near the CDW
quantum critical point.
Finally, the absence of spontaneous magnetic fields below the onset of
superconductivity, as inferred
from zero-field $\mu$SR measurements, confirms that time-reversal
symmetry is preserved in the superconducting state of
CuIr$_{2-x}$Ru$_x$Te$_4$.

\vspace{1pt}
\begin{acknowledgments}
This work was supported by the Natural Science Foundation of Shanghai
(Grant Nos.\ 21ZR1420500 and 21JC140\-2300), the Fundamental Research Funds for the Central Universities, and the Schweizerische 
Nationalfonds zur F\"{o}rderung der Wis\-sen\-schaft\-lichen For\-schung 
(SNF) (Grant Nos.\ 200021\_188706 and 206021\_139082). 
H.Q.Y.\ acknowledges support from the National Key R\&D Program of
China (No.\ 2017YFA0303100 and No.\ 2016YFA0300202), the Key R\&D Program
of Zhejiang Province, China (No.\ 2021\-C01002), the National Natural Science
Foundation of China (No.\ 11974306). 
\end{acknowledgments}

\appendix
\section{\label{appendix} TF-80\,mT $\mu$SR spectra}

\begin{figure}[!thp]
	\centering
	\includegraphics[width=0.5\textwidth,angle= 0]{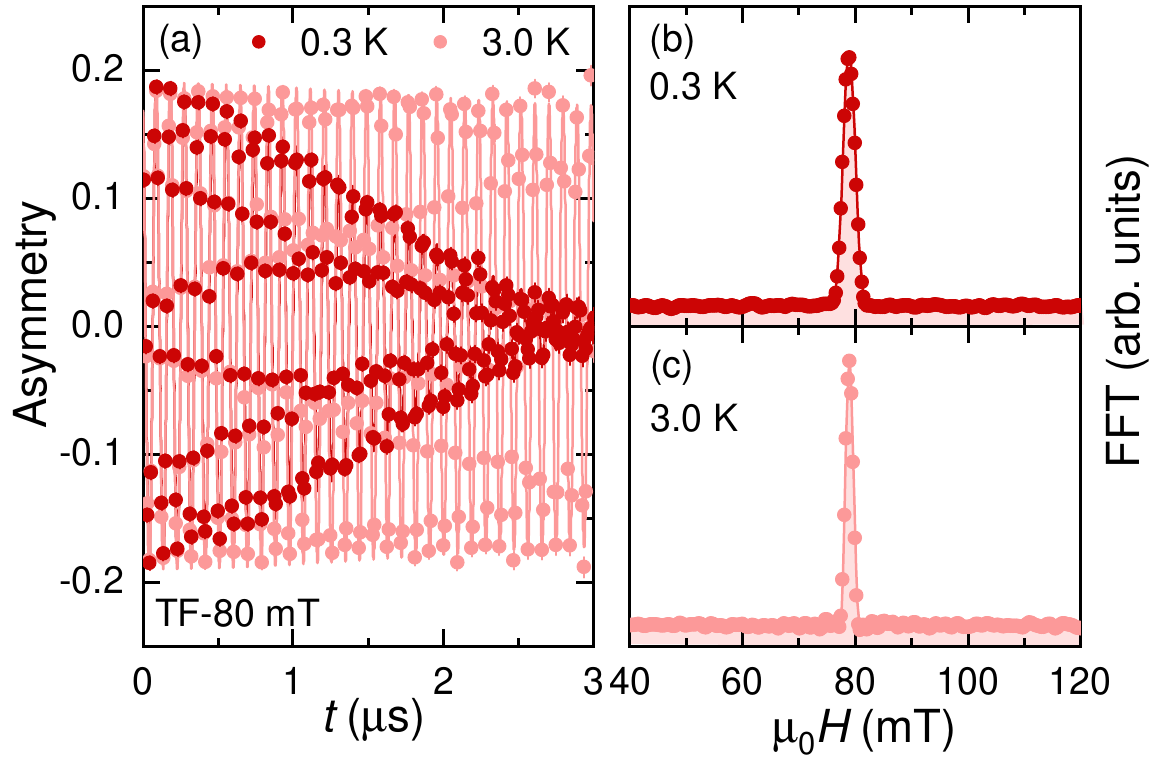}
	\caption{\label{fig:TF-80mT} (a) TF-$\mu$SR spectra collected in an applied field of 80\,mT in both the superconducting- and normal states for CuIr$_2$Te$_4$. 
		The respective real part of the Fourier transforms of TF-80\,mT $\mu$SR spectra are shown in (b) and (c) for 0.3\,K and 3.0\,K, respectively. 
	    Solid lines are fits to Eq.~\eqref{eq:TF_muSR} using a single oscillation.}
\end{figure}

\bibliography{CuIrTe.bib}

\end{document}